\documentclass{pramana}

\usepackage{graphicx,amsmath,bm}

\begin{document}

%%paper title
%%For line breaks \\ can be used within title 
\title{Saha Ionization and Pair production in Rainbow Rindler Metric}

%%author names are separated by comma (,) 
%%use \and before the last author name 
%%\textsuperscript{number} is used for affiliation
%%use a * along with the number separated by comma
%% for the  author for correspondence

\author{Vishnu S Namboothiri\textsuperscript{1}, Gautham Varma K\textsuperscript{1}}
\affilOne{\textsuperscript{1} Center for Science in Society, Cochin University(CUSAT), Kochi, Kerala\\
\affilTwo{\textsuperscript{2} Government College, Kottayam, Kerala}}

%%escape two column mode for title, affiliation and abstract
%%by giving \twocolumn command as shown

\twocolumn[{

\maketitle

%%include \corres to print the corresponding author Email id
\corres{\textsuperscript{1}ramharisindhu@gmail.com,\textsuperscript{2}gauthamvarma31@gmail.com}

%%include \msinfo for
%%manuscript information such as
%%received, revised and accepted dates
%%

%%abstract
\begin{abstract}
The energy of a particle in the Rainbow Rindler metric is derived using Hamilton's variational principle. Saha Ionization equation and Pair production in Rainbow Rindler metric have been investigated. Saha ionization equation and Pair production in Rainbow Rindler metric depend on corrected acceleration which is energy-dependent. This means energy ($E$) and metric tensor ($g_{\mu \nu}(E)$) depends on the inertial observer. 
 The physical implications of energy-dependent acceleration, observer-dependent(inertial) energy and metric tensor are discussed. On comparing the above results with that of Rindler metric we get the result obtained by Sanchari De and Somenath Chakarabarthy by applying the correspondence principle.
\end{abstract}

%%insert keywords separated by comma using \keywords{words}

%%include \pacs{number} to print the PACS number
}]
%%close the twocolumn escape here

%%include \doinum{number}for the DOI number in the header
%%include \volnum{number} for the volume number in the header
%%include \year{yyyy} for  year of publication in the header
%%include \pgrange{num--num} page range of article in the header
%%include \artcitid{num} for the article citation id
%%include \lp to print last page of the article
%%include \setcounter{page}{pagenum} for the exact starting page of the article

\doinum{}
\artcitid{}
\volnum{}
\year{}
\pgrange{}
\setcounter{page}{23}
\lp{25}

 \section{Introduction}
It has been proposed by Lorentz, Poincare\cite{1} and A. Einstein \cite{2} that inertial frames can be related by a set of equations called Lorentz transformations in which Maxwell's equations are invariant. Later, the principle of equivalence includes uniformly accelerated frames into the category of inertial frames. As a result, a similar transformation could be applied to uniformly accelerated frames within the context of special relativity.
 \par The local geometry we get as a result of the transformations is called Rindler geometry. For example let us consider two frames $S$ and $S^{\prime}$. $S$ frame is an inertial frame in the presence of a weak gravitational field (that may be produced by a black hole) and $S^{\prime}$ undergoes uniform acceleration with respect to $S$. Then according to the principle of equivalence, it is impossible to distinguish between the frames. The gravitational field is approximated as a constant value within a small spatial region at distance $x$ from the centre of the gravitating object. This is called the local acceleration of the frame. This frame commonly referred to as the Rindler frame has attracted much of the physics community.
 \par Blackhole space times reduce to the Rindler form in the near horizon limit. Thus the Rindler metric acts as a tool to explore many features of black holes. Many interesting phenomena could be explained easily by working on the Rindler metric. These include Unruh effect, pair production, Schwinger effect, Hawking radiation, Bose-Einstein condensation and so on. 
 \par Rindler metric has been widely discussed in the context of classical gravity. However it is widely believed that the classical theory could possibly be an approximation of a quantum theory of gravity in the limit $\hbar \to 0$.
There are several candidates that attempt to unify quantum mechanics and general relativity.
Loop quantum gravity\cite{3,4}, String theory\cite{4}\cite{5}, Non-commutative geometry
\cite{6,7,8,9} are a few among them. Whatever be the nature of the gravity theory it is widely believed that the Planck energy $E_{Pl}= \sqrt{\hbar c^5/G}$ acts as a threshold energy that separates the classical description from the quantum description. If we probe beyond this limit we get a completely new picture.
\par Leading order effects of $E/E_{Pl}$ (E is the energy observed by an inertial observer) around the classical limit has been studied in astronomical and cosmological observations. Tests of thresholds for high energy cosmic rays\cite{10,11,12,13,14,15,16} hint at a possible energy dependence of the speed of light. The same could also be observed in gamma-ray bursts\cite{17} and TeV photons \cite{18}, as well as in tests involving synchrotron radiation\cite{19,20,21} and nuclear physics \\experiments\cite{22}. Related effects may also be detectable shortly in CMB observations\cite{23}. These observations support the modification of the energy-momentum relation in special relativity as follows:
\begin{equation}
E^2 = p^2 +
m^2 + \alpha l_{Pl} E^3 + ...
\end{equation}
$\alpha$ is a
dimensionless constant of
order unity(We put constants: $\hbar=1=c$). Modifications to the energy-momentum relation demand changes in Lorentz symmetry in the classical limit of quantum gravity. Some physicists argue that the Lorentz symmetry should be broken to accommodate the leading order effects into the energy-momentum relation in special relativity. However Deformed special relativity (DSR) or Non-linear special relativity assumes the Lorentz symmetry. Interestingly the transformations act on momentum space non-linearly thereby assuming the leading order terms of $E/E_{Pl}$. The deformed special theory has some interesting consequences.
\par The energy-momentum conservation take place non-linearly in inertial frames. $E_{Pl}$ is invariant for all inertial observers. The invariance of the Planck energy says that the classical regime ($E>E_{Pl}$) and the quantum regime ($E<E_{Pl}$) will be the same for all inertial observers. In other words what is classical in one frame will be classical in all frames. The same applies to what is quantum.Thus, classical and quantum regimes are very precise in this theory.
\par Deformed special relativity is one among a class of theories that are based on a modified principle of special relativity.
\begin{enumerate}
\item{} It is based on the principle of relativity
\item{} In the limit $E/E_{Pl}\to 0$ the speed of photon becomes a universal constant($c$), which is same for all inertial observers.
\item{} Planck energy ($E_{Pl}$) is a universal constant and is same for all inertial observers.
\end{enumerate}
\par Based on these postulates the energy and momentum relation modifies to
\begin{equation}
E^2 f^2 {\left(E/ E_{Pl}\right)} - p \cdot p g^2{\left(E/
E_{Pl}\right)} = m^2
\end{equation}
This equation can be realized as an action of a non-linear map from momentum space to itself.
It is denoted by $U: {\cal P}\rightarrow {\cal P}$ given by
\begin{equation}
U \cdot \left(E , p_i\right) = \left(U_0,U_i\right)= {\left(f{\left(E/E_{Pl}\right)}E
\right)} E 
,g{\left(E/ E_{Pl}\right)}p_i)
\end{equation}

This implies momentum space has a non-linear norm defined by
\begin{equation}
\tilde{L}_a^b = U^{-1} \cdot L_a^b \cdot U
\end{equation}

This norm is preserved under non-linear Lorentz group, given by
\begin{equation}
\tilde{L}_a^b = U^{-1} \cdot L_a^b \cdot U
\end{equation}
where $ L_a^b$ are the generators
\cite{24}.
\par Deformed special relativity has been formulated non-linearly in momentum space. One might ask the question of how to incorporate a theory of gravity that is consistent with deformed special relativity(DSR). Since momentum transformation rules are no longer linear it is nontrivial to merely extend DSR to formulate a consistent theory of gravity.
The possible answers to the question are non-commutative geometry and $\kappa$-Minkowski space-time\cite{7}.
\par However Lee Smolin and João Magueijo suggest that it is pointless to search for a theory that identify dual space representing position and proceed. Instead they propose classical space-time upto leading order in $l_{Pl}$ represented by a one-parameter family of metrics,\:parameterized by the ratio of $E/E_{Pl}$. Thus, the energy of a particle moving in space-time affects the background metric. The modifications to it depend upon the ratio of the energy of the particle and  the Planck energy. The theory of gravity based on these modifications is known as rainbow gravity. The corresponding metric which could be obtained is called Rainbow Rindler metric.
\par This paper is devoted to studying how the Saha equation and pair production takes place in the Rainbow Rindler background.
To investigate the Saha equation in a uniformly accelerated frame or Rindler frame in rainbow gravity, we obtain the results using the principle of equivalence. The Lagrangian of a particle is derived from Hamilton's principle and from that we get the Hamiltonian of a particle(It may be an ion, hydrogen atom or an electron) from standard relations of classical mechanics. The Saha equation and pair production in the Rindler frame have been studied by Sanchari De, Somenath Chakarabarthy. This work extends their work to Rindler metric in Rainbow gravity called Rainbow Rindler metric.
\subsection{Rainbow Gravity}
\begin{enumerate}
\item {}Modified equivalence principle:\\The freely falling observer who measures energies in between $1/R \ll E \ll E_{Pl}$ finds the laws of physics to be, to first order in $1/R$, the same as in modified special relativity.
\item {}Correspondence Principle:\\In the limit of $ E/E_{Pl} \to 0$\ we recover the ordinary general relativity.\cite{26}
\end{enumerate}
\par If Energy of a quantum is $E= hc/\lambda$ Then the restriction $ 1/R \ll E$ implies that $R \gg \lambda $. Thus, we can infer that classical description is only valid at $E \ll E_{Pl}$ above which we expect quantum space-time.
Thus deformed equivalence principle implies that space-time is described by one parameter family of metrics in terms of ortho-normal fields.
\begin{equation}
g(E)=\eta^{\mu \nu}e_{\mu}\left(E\right)\bigotimes e_{\nu}\left(E\right)
\end{equation}
Where the energy dependence of the frame fields is given by $ e_{0}\left(E\right)= \frac{1}{f\left(E/E_{Pl}\right)} \hat{e}_{0}\left(E\right)$ and $ e_{i}\left(E\right)= \frac{1}{g\left(E/E_{Pl}\right)} \hat{e}_{i}\left(E\right)$. When we apply correspondence principle $\left( E/E_{Pl} \to 0  \right)$
\begin{equation}
f\left(E/E_{Pl}\right) \to 1
\end{equation}
\begin{equation}
g\left(E/E_{Pl}\right) \to 1
\end{equation}
Similarly, Einstein's equations can be written as:
\begin{equation}
G_{\mu \nu}=8 \pi G(E) T_{\mu \nu}(E)+g_{\mu \nu}\Lambda\left(E\right)
\end{equation}
Where $G$, $\Lambda$ will depend upon $E$.
Here $E$ is not the energy of space-time but the scale in which one probe the geometry of space-time. If a freely falling observer uses the motion of a collection of particles to probe the geometry of space-time, E is the total energy of the system of particles as measured by that observer\cite{29}.
\subsection{Energy of the particle in Rainbow Rindler frame}
We have the Rainbow Rindler metric as
\begin{equation}
dS^2=c^2\exp\left(\frac{2Agx}{fc^2}\right)\left(dt^2-dx^2\right)-dy^2-dz^2
\end{equation}
\cite{30}.
To calculate Lagrangian of a particle in the Rainbow Rindler frame.
\begin{equation}
S= -\alpha_{0}\int ds= \int L dt
\end{equation}
Where $\alpha_{0}=m_{0}c$.
\begin{equation}
L=-m_{0}c^2\sqrt{\exp\left(\frac{2Agx}{fc^2}\right)\left(1-\frac{v_{x}^2-v_{yz}^2}{c^2}\right)}
\end{equation}
\begin{equation}
v_{yz}^2=v_{y}^2+v_{z}^2
\end{equation}
where v is the velocity of the particle
Then using
\begin{equation}
H=\Sigma_{i} p_{i}v_{i}-L
\end{equation}
It is found that the Hamiltonian of the particle is:
\begin{equation}
H=m_{0}c^2 \exp\left(\frac{Agx}{fc^2}\right) \sqrt{1-\left(\frac{v_{x}^2+v_{yz}^2}{c^2}\right)}
\end{equation}
Since we can write $p=mv$ and $p^2=p_{x}^2+p^2_{yz}$ we can rewrite it as
\begin{equation}
H=m_{0}c^2 \exp\left(\frac{Agx}{fc^2}\right) \sqrt{1-\left(\frac{p^2}{m^2_{0}c^2}\right)}
\end{equation}
\section{ Saha Ionization equation in Rainbow Rindler frame}
We consider a partially ionized hydrogen plasma.
The system has a cylindrical symmetry and the plasma expands with an acceleration A along positive x direction. This x direction is also the symmetry axis of the cylinder. According to the deformed equivalence principle we can say that the freely falling observer that measures particles of energies in between $1/R \ll E \ll E_{Pl}$ finds the laws of physics to be, to first order in $1/R$, the same as in modified special relativity. This together with dynamic equilibrium of reaction allow as to write Saha ionization equation for accelerated observers in Rainbow Rindler frame as:
\begin{equation}
H_{n}+\gamma<-->H^{+}+e^{-}
\end{equation}
along with,
\begin{equation}
\mu(H_n)=\mu(H^+)+\mu(e)
\end{equation}
Here n=1 and under equilibrium condition chemical potential are also equal. Since the number photons are also not conserved chemical potentials for photons are zero.
\begin{equation}
H=m_{0}c^2 \exp\left(\frac{Agx}{fc^2}\right)-\left(\frac{p^2}{2m_{0}\exp\left(-\frac{Agx}{fc^2}\right)}\right)
\end{equation}
We can define $m"=m_{0}\exp\left(\frac{Agx}{fc^2}\right)$ and\\ $m'=m_{0}\exp\left(-\frac{Agx}{fc^2}\right)$.
\begin{equation}
n=\frac{N}{\Delta V}=\frac{4\pi g_d}{h^3}\int_0^\infty p^2dp
\exp\left(-\frac{1}{kT}\left(\frac{p^2}{2m^\prime}+
m\prime\prime c^2-\mu \right)\right)
\end{equation}
Where $g_{d}$ is the degeneracy of species and $\Delta V=A_{c}\Delta x$ is
a small volume element, $\Delta x$ is a small length element in the
$x$-direction at a distance $x$ from the origin
and $A_{c}$ the cross-sectional area of the cylinder(Since we assumed cylindrical symmetry).
The length $\Delta x$ is such that  the deformed principle of equivalence holds.
On evaluating the above integral (It is a Gaussian function of type $ax^2+b$) and rearranging, we get:
\begin{equation}
\mu=m^{\prime\prime}c^2 -kT\ln\left (\frac{g_{d} n_{Q}}{n}\right )
\end{equation}
Where $n_{Q}$ is defined as
\begin{equation}
n_Q=\left (\frac{2\pi m^\prime kT}{h^2}\right )^{3/2}
\end{equation}
is called the quantum concentration for the particular species in the
mixture. we can define,
\begin{equation}
\frac{n(H^+)n(e)}{n(H_n)}=\frac{n_{Q_e}}{g_n} \exp \left (-
\frac{\Delta E_n}{kT}\right ) =R_{A>0} ~{\rm{(say)}}
\end{equation}
Then taking ratio with degeneracy and without degeneracy
\begin{equation}
\frac{R_{A>0}}{R_{A=0}}=\left (\exp\left(\frac{Agx}{fc^2}\right) \right )^{-3/2}
\exp\left (-\frac{Ax\Delta m}{kT}\right )
\end{equation}
Here
\begin{equation}
\Delta E_n=\Delta m^{\prime\prime}c^2= \left (\exp \left(\frac{Agx}{fc^2}\right) \right) \Delta mc^2
\end{equation}
With
\begin{equation}
\Delta m=m\left(H_n\right)-m\left(H^+\right)-m\left(e\right)
\end{equation}
\section{Pair Production}
In this section we consider the electron-positron pair creation at a high
temperature in a local rest frame in the presence of a constant
gravitational field $g$. With the condition
$kT>2m_{0}$ Where $T$ is the temperature.
\begin{equation}
\gamma+\gamma \leftrightarrow e^- +e^+
\end{equation}
we assume that the
interacting electron-positron plasma
are in thermodynamic equilibrium with the photons.
The chemical equilibrium condition is given by
\begin{equation}
\mu(e^+)+ \mu(e^-)=0
\end{equation}
where $\mu(\gamma)=0$ and
\begin{eqnarray}
\mu(e^-)&=&m^{\prime\prime}(e)c^2-kT\ln \left ( \frac{g_en_Q}{n_{e^-}}\right )
~{\rm{and}}~ \nonumber \\
\mu(e^+)&=&m^{\prime\prime}(e)c^2-kT\ln \left ( \frac{g_en_Q}{n_{e^+}}\right )
\end{eqnarray}
we have
\begin{equation}
n_{e^-}n_{e^+}=g_e^2n_Q^2\exp\left( -\frac{2m^{\prime\prime}(e) c^2}{kT}
\right )
\end{equation}
where
\begin{equation}
n_Q=\left ( \frac{2\pi m^\prime(e) kT}{h^2}\right )^{3/2}
\end{equation}
is the quantum concentration for electron and positron.
)
with $m"_e$ replaced by $m_{0}$, the electron rest mass.
If the system is assumed to be neutral then $n_{e^-}=n_{e^+}$.
Then electron-positron concentration with $A>0$ and $A=0$, which is given by
\begin{equation}
\frac{{n_{e^-}n_{e^+}}_{A>0}} {{n_{e^-}n_{e^+}}_{A=0}}
=\left(\exp\left(\frac{Agx}{fc^2}\right) \right )^{-3} \exp\left ( -\frac{2Ax m_{\left(0\right)}}
{kT}\right )
\end{equation}
\section{Physical Significance and Conclusion}
On comparing the above results with that of Rindler metric we get the result obtained by Sanchari De and Somenath Chakarabarthy by applying the limit $E/E_{Pl}\to 0$ (correspondence principle). In the Rindler metric the Saha Ionization equation is given by,
with $A>0$ and $A=0$ can be written as
\begin{equation}
\frac{R_{A>0}}{R_{A=0}}=\left ( 1+\frac{Ax}{c^2}\right )^{-3/2}
\exp\left (-\frac{Ax\Delta m}{kT}\right )
\end{equation}
\cite{31}.
Our result will reproduce the above equation under the correspondence principle and by the Taylor expansion of the exponential function up to first order.
\begin{equation}
\frac{R_{A>0}}{R_{A=0}}=\left (\exp \left(\frac{Agx}{fc^2}\right) \right )^{-3/2}
\exp\left (-\frac{Ax\Delta m}{kT}\right )
\end{equation}
By the Taylor expansion of $\exp\left(\frac{Agx}{fc^2}\right)$  we get, $\exp(\frac{Agx}{fc^2})=1+\frac{Agx}{fc^2}+\frac{A^2g^2x^2}{4f^2c^4}$. Neglect terms containing $c^4$ and we get:
\begin{equation}
\exp\left(\frac{Agx}{fc^2}\right)=1+\frac{Agx}{fc^2}
\end{equation}
Thus, Saha equation in Rainbow Rindler gravity becomes:
\begin{equation}
\frac{R_{A>0}}{R_{A=0}}=\left (1+\frac{Agx}{fc^2} \right )^{-3/2}
\exp\left (-\frac{Ax\Delta m}{kT}\right )    
\end{equation}
electron-positron concentration with $A>0$ and $A=0$ in Rainbow Rindler gravity, which is given by
\begin{equation}
\frac{R_{A>0}}{R_{A=0}}=\left (1+\frac{Agx}{fc^2} \right )^{-3}\exp\left (-\frac{Ax\Delta m}{kT}\right )  \end{equation}

This shows that there is a corrected acceleration which is energy-dependent in Rainbow gravity theory. This means E and $g_{\mu \nu}(E)$ depends on the observer.  If different observers measure different energies for a given process in the Rainbow Rindler frame, then the corresponding metrics will also be different. It is also possible that the same observer assigns E to a particular process then he measures a different Energy $E^{\prime}$ to another process at the same place. In case of geometry also, not only different observers may see a given particle being affected by different metrics, but the same observer may assign different metrics to different particles moving in the same region at the same 
time. This is due to covariance, once we permit for a non-linear representation of the Lorentz group in momentum space.

\begin{equation}
 ds^2=-\frac{\left(dx_{0}^2\right)}{f^2}+\frac{\left(dx_{i}^2\right)}{g^2} 
\end{equation}
This is energy-dependent quadratic invariant \cite{29}. 
\par The energy-dependent acceleration and geometry differ from that of General relativity. So we hope that this result will help to explore the nature of accelerated frames in the Rainbow gravity theory whose physical characteristics depend upon the energy that can be used to probe the nature of space-time. Bose-Einstein condensation and properties of black holes in Rainbow gravity are the further problems that can be investigated. 

\section*{Acknowledgement}
We are very much thankful to Prof Somenath Chakrabarty at Visva-Bharati University for comments on this work. We always remember the help and motivation of Prof Ramesh Babu T(CUSAT),
Prof V.P.N.Nampooiri(CUSAT).

%%References section

\end{document}